\documentstyle[a4]{article}

\begin{document}

\newcommand{\er}{\end{eqnarray}}
\newcommand{\br}{\begin{eqnarray}}
\newcommand{\be}{\begin{equation}}
\newcommand{\ee}{\end{equation}}
\newcommand{\epe}{\end{equation}}
\newcommand{\bea}{\begin{eqnarray}}
\newcommand{\eea}{\end{eqnarray}}
\newcommand{\ba}{\begin{eqnarray}}
\newcommand{\ea}{\end{eqnarray}}
\newcommand{\epa}{\end{eqnarray}}
\newcommand{\ar}{\rightarrow}
\newcommand{\dslash}{\partial\!\!\!/}
\newcommand{\aslash}{a\!\!\!/}
\newcommand{\bslash}{b\!\!\!/}
\newcommand{\kslash}{\kappa\!\!\!/}
\newcommand{\fslash}{f\!\!\!/}
\newcommand{\Dslash}{D\!\!\!\!/}

\def\r{\rho}
\def\D{\Delta}
\def\R{I\!\!R}
\def\l{\lambda}
\def\D{\Delta}
\def\d{\delta}
\def\T{\tilde{T}}
\def\k{\kappa}
\def\t{\tau}
\def\f{\phi}
\def\p{\psi}
\def\z{\zeta}
\def\ep{\epsilon}
\def\hx{\widehat{\xi}}
\def\na{\nabla}

\begin{center}

{\bf Topologically massive gauge theories from first order theories in
arbitrary dimensions. }

\vspace{1.3cm} M. Botta Cantcheff\footnote{e-mail: botta@cbpf.br}

\vspace{3mm} Centro Brasileiro de Pesquisas Fisicas (CBPF)

Departamento de Teoria de Campos e Particulas (DCP)

Rua Dr. Xavier Sigaud, 150 - Urca

22290-180 - Rio de Janeiro - RJ - Brasil.

\end{center}

\begin{abstract} 

We thereby prove that a large class of topologically massive theories of the Cremmer-Scherk-Kalb-Ramond-type in any $d$ dimensions corresponds to gauge non-invariant first-order theories that can be interpreted as self-dual models.
\end{abstract}

\vspace{1cm}

The apparent clash between gauge 
symmetry and massive gauge bosons is avoided in the framework
of topologically massive gauge theories, as it is the case for the well-known 
Maxwell-Chern-Simons \cite{Deser}
and Cremmer-Scherk-Kalb-Ramond models (CSKR) \cite{Cremer,Kalb,Lahiri1,bf}. They
 illustrate how Abelian gauge bosons may
 be attributed a physical mass without the need of bringing about Higgs scalars and
 spontaeous symmetry breaking. This a fundamental motivation to study this type of theories 
in different space-time dimensions.

This paper has a two-fold purpose:
to construct first-order formulations of 
topologically massive theories which involve BF-terms (topological coupling between different gauge forms
 \cite{Cremer,bf} ) in arbitrary dimensions
and for all possible tensorial ranks; and afterwards, to argue that,
 by considering doublets of field-forms \cite{dob}, 
 these first order (gauge non-invariant) formulations constitute self-dual models, 
 close in spirit to the Self-Dual system in $(2+1)$-dimensions
first introduced by Townsend, Pilch and van Nieuwenhuizen \cite{TPvN}.

There are some recent works \cite{HS, HS2, HS3} pointing out that Cremmer-Sherk-Kalb-Ramond models in dimension four, 
which include in their Lagrangian BF-terms
are dual equivalent to first order ones. These authors employed
the Hamiltonian embedding procedure by Batalin, Fradkin and Tyutin \cite{bft}.
Dualization of these models has also been studied
by Smailagic and Spallucci \cite{spal}, coming to results different from those found in this
letter.

The paralell between these first order BF-theories, at any space-time dimension, 
and the Self-Dual (SD) theories in $(2+1)$, exploited in this work, has recently been
 pointed out by Harikumar et al \cite{HS} in the case 4-dimensional case;
however, they mention a difficulty in establishing this connection as due to the
the imposibility of defining self-duality in dimensions that are not of the form
 $d=4k-1$ ($k\in Z_+$). Here, this objetion is 
by-passed from the very starting point, by defining the dual operation on the space of pairs
of gauge forms.

 In the present approach, we proceed further and use this paralellism to define
a SD model in arbitrary space-time dimension, and 
adapt the proof proposed by Deser and Jackiw in $2+1$-d \cite{DJ} to
  manifestly show the dual
  correspondence between generic topologically massive models (CSKR)
and the already mentioned SD theories in d-dimensions.

Finally, we shall confirm the result
 recently presented in ref. \cite{HS} in four dimensions, as a particular
 case and generalize it to all dimensionalities. 

\vspace{1cm}

First, let us briefly describe the well-known MCS-SD duality in 
 $(2+1)$-dimensions. One currently defines the duality operation by 
\be
\label{1}
\mbox{}^{\star} f_\mu =\frac{\chi}{\mu}\,
\epsilon_{\mu\nu\lambda}\,\partial^\nu f^\lambda\,,
\ee
 where $\mu$ is a mass parameter here introduced to render the
 $\mbox{}^{\star}$-operation
 dimensionless. This is basically a functional curl (rotational operator).

We name {\it self(anti-self)-duality}, when the relations
$\mbox{}^{\star} f = \pm f $ are (respectively) satisfied.

The so-called Self-Dual Model (Townsend, Pilch and van Nieuwenhuizen \cite{TPvN})
 is given by the following action,

\be
\label{180}
 S(f)= \int\, d^3x\:\Bigg(\frac {\chi }{2\mu}\,
\epsilon_{\mu\nu\lambda}\,f^\mu\,\partial^\nu f^\lambda -
\frac{1}{2}\, f_\mu f^\mu \Bigg)\,.
\ee

The equation of motion is the self-duality relation:
\be
\label{190}
f_\mu =\frac{\chi}{\mu}\,
\epsilon_{\mu\nu\lambda}\,\partial^\nu f^\lambda\,.
\ee
This model is claimed to be chiral, and the chiralities $\chi=\pm 1$ result
 defined precisely from this self-duality.

On the other hand, the gauge-invariant combination
of a Chern-Simons term with a Maxwell action, \be \label{370}
S_{MCS}[A]= \int\, d^3x\:\Bigg(\frac{1}{4 \mu^{2}}
F^{\mu\nu}F_{\mu\nu} -
\frac{\chi}{2\mu}\,\epsilon^{\mu\nu\lambda}\,A_{\mu}\,
\partial_{\nu}A_{\lambda}\Bigg)\,,
\ee
is the topologically massive theory, which is known to be equivalent \cite{DJ} to
 the self-dual model (\ref{180}).
$F_{\mu\nu}$ is the usual Maxwell field strength,

\be \label{285} F_{\mu\nu}[A] \equiv \partial_{\mu}A_{\nu} -
\partial_{\nu}A_{\mu}\, = 2\partial_{[\mu}A_{\nu]}. \ee

This equivalence has been verified with the Parent Action Approach \cite{suecos}.
We write down the general parent action proposed by Deser and Jackiw in \cite{DJ}, which proves this equivalence:

 \be
 \label{masterDJ}
 {\cal S}_{Parent}[A,f]= \chi\,{\cal S}_{CS}[A] -\int\,d^{3}x\, \left[ \epsilon^{\mu\nu\lambda}\,
F_{\nu\lambda} [A]
 f_{\mu} \, +  \mu f_{\mu } f^{\mu }\right] ,
\ee

where 
\be
\label{CS}
{\cal S}_{CS}[A] \equiv \int\,d^{3}x\, \epsilon^{\mu\nu\lambda}\, \left( 
{A}_\mu\partial^{}_{\nu}A_{\lambda} \right),
\ee
is the Chern-Simons action \cite{Deser}.

\vspace{0.7cm}

For general dimensions, it is possible to define self (and anti-self)-duality
for pairs (doublets) of form-fields with different ranks \cite{dob}; so,
 a paralell of this structure with the one in $d$ dimensions will be observed.

\vspace{0.7cm}

 The problem of defining the Hodge duality for all
dimensions is well-known; for instance, in Lorentzian four-dimensional spacetime,
 the main obstruction to
self-duality comes from the relation of
double-dualization \footnote{For a  generic $q$-form,  $A$,
the Hodge dual is defined by
\be (\mbox{}^{*} A )^{\mu_{q+1}\cdots \mu_{d}} = \frac 1{q!} \;
\epsilon^{\mu_1\cdots \mu_{d}} A_{\mu_{1}\cdots \mu_{q}}.\ee}  for a rank-two tensor:

\be \label{Q10} \mbox{}^{**} F = (-1)^{s}\; F ,\ee 
where $s$ is the
signature of space time
\footnote{i.e, 
this is the number of minuses occurring in the metric.}.  For the case
 of the Lorentzian metric, where $s$ is an
odd number, the self-duality concept seems inconsistent with the
double dualization operation due to the minus sign in (\ref{Q10}).
This problem remains for dimensionality $d=4m$ ($m\in Z_+$)
\cite{kn:DGHT}; in contrast, it is absent for $d=4m-2$.
Thus, self-duality is claimed to be well-defined (only) in such a
dimensionality. 
First, let us recall that (\ref{Q10}) has led to the prejudice
that the (Abelian) Maxwell theory would not possess manifest
self-duality solutions.

 The resolution of this obstruction came with
the recognition of an internal two-dimensional structure hidden in
the space of fields. Transformations in this internal duality
space extends the self-duality concept to this case and is
currently known under the names of Schwarz and Sen\cite{SS}, but
this deep unifying concept has also been appreciated by
others\cite{Z}.  The actions worked out correspond to self-dual
and anti-self-dual representation of a given theory and make use
of the internal space concept.  The duality operation is now
defined to include the internal (two dimensional) index $(i\,,\,j )$ in the
fashion

\be
\label{Q20} \hat F^i =e^{ij}\; \mbox{}^{*}F^j \ee

where the $2 \times 2$-matrix, $e$, depends on the signature and dimension 
of the spacetime in the form:

\be
\label{a20}
e^{\alpha\beta}= \cases{\sigma_1^{\alpha\beta},&if $d=4m-2$\cr
	\ep^{\alpha\beta},&if $d=4m$\, ,\cr}
\ee

with $\sigma_1^{\alpha\beta}$ being the first of the Pauli matrices
and $\epsilon^{\alpha\beta}$ is the totally antisymmetric $2 \times 2$ matrix with $\ep^{1,2}=1$.
 
 The double dualisation operation,
\be
\label{Q30} ( \hat{\hat F} )_{}^i  \,= \, F^i \ee generalizes (\ref{Q10}) to allow
consistency with self-duality. It has been shown that this prescription
works in the construction of self-dual Maxwell actions \cite{RW} .

\vspace{0.7cm}

This structure has always been considered in the literature only for
tensorial objects where the field has the same tensorial rank that
its corresponding dual.
Howerver, we may generalize further these ideas, introducing
more general doublets \cite{dob}. 

\vspace{0.7cm}

Let a $d$-dimensional space-time with signature $s$, and a generic
element $\Phi \equiv (a\,,\,b) $ in the space  $H_p \equiv \Lambda_p \times \Lambda_{d-p}$.
i.e, $a\, , \,b$ are either a $p$-form and a $(d-p)$-form respectively. Thus, one may define 
a {\it Hodge-type} operation for these
 objects\footnote{Which clearly includes the case $p=d/2$ described above.}
 by means of
\be
\mbox{}^{*} \Phi \equiv ( \mbox{}^{*} b  \, , \,S_p \,\mbox{}^{*} a ),
\ee

where $S_q$ is a number defined by
 the double dualisation operation, for a generic $q$-form $A$:
$\mbox{}^{*}(\mbox{}^{*}A) = S_q \, A \,.$
This depends on the signature ($s$) and dimension of the spacetime in the form
$S_q =(-1)^{s+q[d-q]}$.

Notice that $\mbox{}^{*}$ applied to doublets is defined such that
 its components are interchanged
 with a supplementary change of sign for the second component.

Notice that this Hodge-type self (anti-self)-duality {\it is well-defined},
since 
 \be
 \mbox{}^{*} \Phi =\pm\,  \Phi \, ,
\ee
is consistent with the double dualization requirement, $\mbox{}^{*}(\, \mbox{}^{*}\, \Phi  )=\, \Phi $.

\vspace{0.7cm}

For our purpose in this paper, we are more interested in proposing 
and working with another type of dual-operation
 of a similar nature to the duality we describe above for the case of $2+1$-dimensions.

Let a $d$-dimensional space-time with signature $s$: we consider the tensor doublet,
\be
{\cal F} :=(f_{\mu_1 \cdots\mu_p} ,  g_{\mu_1 \cdots\mu_{d-p-1}} ), \ee
 where $f$ is 
a $p(<d)$-form ( a totally antisymmetric tensor type $(0;p)$ ) , and $g$ is a $(d-p-1)$-form.
 ${\cal F} $ is an element of 
the space $\Delta_p \equiv \Lambda_p \times \Lambda_{d-[p+1]}$.

There is also a well defined notion of self (and anti-self)-duality for the objects in this space.
 Consider the action with topological coupling:
\bea
S_{DSD}[{\cal F}] \equiv \int dx^d \,[{-2\over m}\left(g_{\mu_1\cdots\mu_{d-p-1}} 
\epsilon^{\mu_1\cdots \mu_{d}} \partial_{\mu_{d-p}} f_{\mu_{d-p+1}\cdots\mu_{d}}\right) 
+ \, [p+1]! \, g_{\mu_1 \cdots\mu_{d-p-1}}g^{\mu_1 \cdots\mu_{d-p-1}}+ \nonumber\\
+ (-1)^s [d-p-1])!\,f_{\mu_1 \cdots\mu_p}f^{\mu_1 \cdots\mu_p} ].
\label{DSD}
\eea

For a more concise notation, in terms of forms, consider the following definitions:
 $d(f , g) \equiv (df\, , dg)$, and 
\be
\mbox{}^{*}\, (df \, , ~ dg) \equiv (\mbox{}^{*}dg \, ,~ (-1)^{p+1}S_{p+1} ~\mbox{}^{*}df)\,,
\ee
once more, the operation $\mbox{}^{*}$ applied to objects 
in $\Delta_p$ supposes components interchange 
 and an appropriate modification of sign for the second component.

In so doing, the equations of motion derived from the action (\ref{DSD})
read as
\be
\label{sdrel}
{\cal F} = {1 \over  m }\,\mbox{}^{*} d{\cal F}   ,
\ee
where $m$ is a mass parameter introduced for dimensional
 reasons. It may trivially be verified that these equations require
 that ${\cal F}$ satisfies a Proca equation with mass $m$
\footnote{We are considering here a mass parameter $m$,
for simplicity; however, it could be replaced by a diagonal matrix
in a more general fashion. }

Notice that the equation (\ref{sdrel}) looks like (\ref{190}).
In that sense, we state that $S_{DSD}$
describes {\it doublet}-self-duality.

The other remarkable similarity of this model with SD (in $(2+1)$-d)
is that this is dual to a topologically massive theory (CSKR-type,
with BF-coupling between two gauge forms) in the same way
that  the SD-MCS duality in three dimensions. 
This constitutes our main point, which confirm and generalize some recents results \cite{HS}.
Below, we are going to prove this correspondence.

Note that this structure is insensitive to the space-time dimensions and 
the tensorial ranks of the
doublet components. Thus, a Deser-Jackiw-inspired parent action may be written in $d$
 space-time dimensions.

Consider the doublet of gauge fields $ {\cal A} \equiv (a_{\mu_1 \cdots\mu_p} ,  b_{\mu_1 \cdots\mu_{d-p-1}} )$
in addition to $ {\cal F} =(f_{\mu_1 \cdots\mu_p} ,  g_{\mu_1 \cdots\mu_{d-p-1}} ) $;
the parent action proposed is:

\bea
{\cal S}_{P}[{\cal A} , {\cal F}] = {\cal S}_{BF}[{\cal A} ] - \int dx^d \, \epsilon^{\mu_1\cdots \mu_{d}} 
\left[ b_{\mu_1\cdots\mu_{d-p-1}} \partial_{\mu_{d-p}} f_{\mu_{d-p+1}\cdots\mu_{d}} 
+ g_{\mu_1 \cdots \mu_{d-p-1}} \partial_{\mu_{d-p}} a_{\mu_{d-p+1}\cdots\mu_{d}}\right] + \nonumber\\
+ \int dx^d \,{m\over 2}\left( [p+1]! \, g_{\mu_1 \cdots\mu_{d-p-1}}g^{\mu_1 \cdots\mu_{d-p-1}}
+ (-1)^s [d-p-1]!\,f_{\mu_1 \cdots\mu_p}f^{\mu_1 \cdots\mu_p}\right),
\label{masterDJdob}
\eea
where 
\be
\label{BF}
{\cal S}_{BF}[{\cal A}] \equiv \int dx^d \,\left[- b_{\mu_1\cdots\mu_{d-p-1}} 
\epsilon^{\mu_1\cdots \mu_{d}} \partial_{\mu_{d-p}} a_{\mu_{d-p+1}\cdots\mu_{d}} \right]
\ee
may be recognized as a BF-action.

Varying ${\cal S}_{P}$ with respect to ${\cal F}$, we obtain
 \be
{\cal F}= - {1 \over  m }\,\mbox{}^{*} d{\cal A};
 \ee
plugging this back into (\ref{masterDJdob}), we recover the topologically massive
gauge action (CSKR):
\be
\label{GMCS}
 {\cal S}_{CSKR}[{\cal A}]= {\cal S}_{BF}[{\cal A}]  - \int \frac{d^{d}x}{2m} \,
\left( (-1)^s \,[d-p-1]! \,(\partial_{[\mu}a_{\mu_1 \cdots\mu_p ]})^2 + \,[p+1]!\,
(\partial_{[\mu}b_{\mu_1 \cdots\mu_{d-p-1}]})^2  \right).
\ee

We shall observe that this is invariant
 under the gauge transformations; $ {\cal A}  \to {\cal A}  +
 d{\cal  D}$, where $d{\cal D}$ is a {\it pure gauge doublet}, i.e, it is a pair of 
exact differentials of $(p-1,d-p-2)$-forms.

Now, we vary ${\cal S}_{P}$ with respect to ${\cal A}$ and obtain:
 \be
\mbox{}^{*}d({\cal A}-{\cal F})=0 ;\ee
or in components,
\ba
\mbox{}^{*}d(a-f)&=&0\nonumber\\
\mbox{}^{*}d(b-g)&=&0\, .
\ea
 This implies that  the differences $a-f$ and $b-g$ may locally be
  written as exact forms; therefore,
 one it is possible to express the solution to these equations as
\be  {\cal A} = {\cal F} + d{\cal D} .\ee Putting this back into the
action (\ref{masterDJdob}) , we recover the SD theory (\ref{DSD}) up to
topological terms.

This completes the proof of our main statement.

\vspace{0.7cm}

 As an example, one can particularize this result for the special dimensionality,
$d=3+1$. In this case, only two tensor doublets may be chosen: ${\cal G}
=(A_\mu , B_{\nu\rho})$ and ${\cal H}=(\phi, F_{\nu\rho\alpha})$.
The first one describes a Cremmer-Scherk-Kalb-Ramond massive spin-one
particle and, by virtue of the general result proven before, 
its dynamics may alternatively be described by either, 
the Cremmer-Scherk-Kalb-Ramond theory
\be
\label{par2}
S_{CSKR}({\cal G}) =\int d^{4}x\left(\frac{1}{2m}\partial_{[ \rho} A_{\mu ]}\partial^{[ \rho} A^{\mu ]} -
\frac{1}{2m} \partial_{[ \rho} B_{\mu \nu ]}\partial^{[ \rho} B^{\mu \nu]} +
B_{\mu\nu}\epsilon^{\rho\mu\nu\sigma}\partial_{\rho}A_{\sigma}\right),
\ee
or the first-order SD model:
\be
\label{dob2k}
S_{DSD}({\tilde {\cal G}})=
 \int d^{4}x \left( - \, {\tilde A}_{\sigma}{\tilde A}^{\sigma} +  {\tilde B}_{\mu\nu}{\tilde B}^{\mu\nu}
+ {1\over m} {\tilde A}_{\sigma}\epsilon^{\sigma\rho\mu\nu}
\partial_{[ \rho}{\tilde B}_{\mu\nu ]}  \right) ,
\ee
which is gauge non-invariant. This confirm the result recently presented in ref. \cite{HS}
\footnote{However, in ref. \cite {HS}, this duality is shown by using the 
Batalin, Fradkin and Tyutin embedding technique \cite{bft}.}.

The second possible doublet in four dimensions
describes a scalar (spin-zero) massive particle whose dynamics may be given
by a topologically massive action,
\be
S_{TM}({\cal H})= \int
d^{4}x\left(\frac{1}{2m} \partial_{[\mu}F_{\nu\rho\alpha]}\partial^{[\mu}F^{\nu\rho\alpha]}
 - 3! \partial_{\mu}\phi ~\partial^{\mu}\phi
+ \phi \epsilon^{\mu\nu\rho\alpha}\partial_{\mu}F_{\nu\rho\alpha}\right),
\ee
or alternatively, by a first order (SD) model :
\be
S_{DSD}( {\tilde {\cal H}})=  \int
d^{4}x\left({\tilde \phi}^2 - \frac{3!}{m} {\tilde F}_{\mu\nu\rho}{\tilde F}^{\mu\nu\rho} + \frac{2}{m} 
{\tilde \phi}\epsilon^{\mu\nu\rho\alpha}\partial_{\mu}{\tilde F}_{\nu\rho\alpha}\right).
\ee

\vspace{0.7cm}

Doublet Hodge Duality has been defined in a similar sense to the duality in $3d$ \cite{dob}.
This suggests a list of formal correspondences between theories
in $3d$ which involve self-duality and similar models in other dimensions. 
This constitutes by itself a very important application of this formalism since one can,
 in principle, translate the constructions of 3d to arbitrary dimensions.

An interesting possibility that we open up 
is the study of bosonization in arbitrary dimensions, mainly in higher dimensions. This is
 not a trivial matter \cite{luscher,marino,banmarino}, but with the help
 of the technique suggested here, $d\geq 4$ bosonization comes out 
in connection with a topologically massive model that mixes different gauge forms. Results
 on this issue shall
soon be reported elsewhere \cite{boson}.

{\bf Aknowledgements}: The author is indebted to J. A. Helayel-Neto 
for invaluable discussions and pertinent corrections on the 
manuscript. Thanks are due to the GFT-UCP
 for the kind hospitality. CNPq is also acknowledged for the invaluable financial help.


\begin{thebibliography}{99}
\bibitem{Deser} S. Deser, R. Jackiw, and S. Templeton, Ann. Phys. {\bf
140} (1982) 372.
\bibitem{Cremer} E. Cremer and J. Scherk, Nucl. Phys. {\bf B72} (1974)
117.
\bibitem{Kalb} M. Kalb and P. Ramond, Phys. Rev. {\bf D9} (1974) 2273.
\bibitem{Lahiri1} T.J. Allen, M.J. Bowick, and A. Lahiri, Mod. Phys.
Lett. {\bf A6} (1991) 559; R. Amorim and J. Barcelos-Neto, Mod. Phys.
Lett. {\bf A10} (1995) 917.

\bibitem{bf} A. Aurulia and Y. Takahashi, Prog. Theor. Phys. {\bf 66} (1981)
693; Phys. Rev. {\bf D23} (1981) 752; T. J. Allen, M. J. Bowick and
A. Lahiri, Mod. Phys. Lett. {\bf A6} (1991) 559.

\bibitem{dob} M. Botta Cantcheff, ''Hodge-type self(antiself)-duality for general
 p-form fields in arbitrary dimensions ", hep-th/0107123.

\bibitem{TPvN} P. K. Townsend, K. Pilch and P. van Nieuwenhuizen,
 Phys. Lett. B 136 (1984) 38.

\bibitem{HS}
E. Harikumar, M. Sivakumar, Nucl Phys-B 565 (2000) 385, and references therein.

\bibitem{HS2}
E. Harikumar, M. Sivakumar, Mod.Phys Lett A 15 (2000) 121.

\bibitem{HS3}
E. Harikumar, M. Sivakumar, ''Hamiltionian vs Lagrangian embedding
of massive spin one theory involving two form field." ,hep-th/0104107.

\bibitem{bft} I. A. Batalin and E. S. Fradkin, Nucl. Phys. {\bf B279}
(1987) 514;
I. A. Batalin and I. V. Tyutin, Int. J. Mod. Phys. {\bf A6} (1991) 3255.


\bibitem{spal} E.Smailagic, E. Spallucci, Phys. Rev. D61 (2000) 067701.

\bibitem{DJ} S. Deser and R. Jackiw, Phys. Lett. B 139 (1984) 2366.

\bibitem{suecos} For a review in the use of the master action
 in proving duality in diverse areas see: S. E. Hjelmeland, U. Lindstr\"om, UIO-PHYS-97-03,
 May 1997. e-Print Archive: hep-th/9705122.

\bibitem{kn:DGHT} S. Deser, A. Gomberoff, M Henneaux and C. Teitelboim, {\it Duality, Self-Duality, 
Sources and Charge Quantization in Abelian $N$-Form Theories}, hep-th/{\bf 9702184};
C. Wotzasek, Phys.Rev.D58:125026,1998; R. Banerjee, C. Wotzasek, Phys.Rev.D63:045005,2001

\bibitem{SS} J. Schwarz and A. Sen, Nucl. Phys. B411 (1994) 35

\bibitem{Z} D. Zwanziger, Phys. Rev. D3 (1971) 880;
 S. Deser and C. Teitelboim, Phys. Rev. D13 (1976) 1592.

\bibitem{RW}R.Banerjee and C.Wotzasek, Nucl.Phys B527(1998) 402.



\bibitem{luscher} M.L\"uscher, Nucl. Phys. B326 (1989) 557.
\bibitem{marino} E. C. Marino, Phys. Lett. B263 (1991) 63.
\bibitem{banmarino} R. Banerjee, C. Marino, Phys. Rev. D56 (1997) 3763.
\bibitem{boson}  M. Botta Cantcheff, J. A. Helayel-Neto, work in progress.



\end{thebibliography}
\end{document}